\documentclass[%
 reprint,
 showkeys,
 amsmath,amssymb,
 aps,
]{revtex4-2}

\usepackage{graphicx}
\usepackage{dcolumn}
\usepackage{bm}

\usepackage{gensymb}
\usepackage{amssymb,amsmath}
\usepackage{siunitx} 
\usepackage{textcomp}
\usepackage{xcolor} 
\graphicspath{ {images/} }

\usepackage{sidecap}
\makeatletter
\def\@fnsymbol#1{\ensuremath{\ifcase#1\or \dagger\or *\or \ddagger\or
   \mathsection\or \mathparagraph\or \|\or **\or \dagger\dagger
   \or \ddagger\ddagger \else\@ctrerr\fi}}
   \makeatother
\usepackage{soul}

\begin{document}

\preprint{APS/123-QED}

\title{Infection percolation: \\A dynamic network model of disease spreading}

\author{Christopher A. Browne}
    \altaffiliation{These authors contributed equally to this work.}
\author{Daniel B. Amchin} 
    \altaffiliation{These authors contributed equally to this work.}
\author{Joanna Schneider}
    \altaffiliation{These authors contributed equally to this work.}
\author{Sujit S. Datta}
    \altaffiliation{To whom correspondence should be addressed:\\ ssdatta@princeton.edu}

\affiliation{Department of Chemical and Biological Engineering, Princeton University, Princeton, NJ 08544}


\date{\today}

\begin{abstract}

\noindent Models of disease spreading are critical for predicting infection growth in a population and evaluating public health policies. However, standard models typically represent the dynamics of disease transmission between individuals using macroscopic parameters that do not accurately represent person-to-person variability. To address this issue, we present a dynamic network model that provides a straightforward way to incorporate both disease transmission dynamics at the individual scale as well as the full spatiotemporal history of infection at the population scale. We find that disease spreads through a social network as a traveling wave of infection, followed by a traveling wave of recovery, with the onset and dynamics of spreading determined by the interplay between disease transmission and recovery. We use these insights to develop a scaling theory that predicts the dynamics of infection for diverse diseases and populations. Furthermore, we show how spatial heterogeneities in susceptibility to infection can either exacerbate or quell the spread of disease, depending on its infectivity. Ultimately, our dynamic network approach provides a simple way to model disease spreading that unifies previous findings and can be generalized to diverse diseases, containment strategies, seasonal conditions, and community structures.
\end{abstract}

\maketitle

\section*{Introduction}

Epidemic spreads---such as the 1918 flu, HIV/AIDS, and COVID-19 pandemics---highlight the critical importance of infectious disease modeling in our everyday lives. Mathematical models can provide key insights into the process by which disease is transmitted between individuals, help to forecast how a disease will continue to spread through a population, and assess the efficacy of different interventions. Developing accurate, computationally-tractable, and generally-applicable models of disease spreading is therefore of critical importance to public health.

The dynamics of infectious disease spreading are often modeled using compartmental models \cite{brauer2019compartment,brauer2012mathematical} that employ the framework of reaction kinetics. For example, in one representation, members of a population are divided into three intermixed groups: those who are susceptible to infection ($S$), currently infected ($I$), or recovered from infection ($R$), with transitions from $S\rightarrow I$ and $I\rightarrow R$ occurring at prescribed rates. This standard model, known as the $SIR$ model or the $SI$ model in the absence of recovery, can successfully predict the initial exponential and eventual logistic dynamics of the total amount of infection in a population for many diseases \cite{grassly2008mathematical,ross1916application1,kermack1927contribution1,okyere2016fractional}. Thus, the $SIR$ model provides a useful approach for disease modeling that is well established.

However, a notable omission of this model is that it does not consider the discrete, spatially-separated interactions between members of a population. Instead, it assumes a well-mixed population for which the crucial dynamics of disease transmission between individuals are lumped into the basic reproduction number $R_{0}$, a macroscopic parameter describing the mean number of secondary transmissions from each infection. In practice, this quantity is used as a fitting parameter---limiting projection capabilities when the interactions between individuals change even slightly \cite{eckalbar2015dynamics,glass2004effect,van2001measles,ball2006optimal,viboud2005multinational,kilbourne2006influenza}. A recent example highlighting this deficiency is the spread of COVID-19 in China: containment policies imposing spatial barriers and suppressing individual interactions are thought to have hindered exponential growth of infection, yet this pivotal effect cannot be captured by the classic $SIR$ model without invoking additional fitting parameters \cite{maier2020effective}.

To address this issue, sophisticated extensions of this model have been developed to explicitly incorporate spatial variations in spreading \cite{bailey1975mathematical,mollison1977spatial,grassberger1983critical,bunde1985universality,herrmann1986geometrical,grenfell2001travelling,may2001infection,moreno2002epidemic,newman2002percolation,newman2002spread,lloyd2005superspreading,miller2007epidemic,kenah2007second,trapman2007analytical,davis2008abundance,lagorio2009effects,parshani2010epidemic,karrer2010message,neri2011effect,neri2011heterogeneity,ochab2011shift,wang2017unification,brauer2012mathematical}. For example, in one approach, different subpopulations---each modeled using different $SIR$ dynamics---are coupled together \cite{brauer2012mathematical,stonesynch,arenascovid,brauerstructure}. While powerful, this approach still employs lumped parameters for each subpopulation that do not explicitly consider differences between discrete individuals. Percolation theory provides another powerful way to explore disease spreading throughout a social network composed of discrete, spatially-separated individuals, yielding insights into the onset of disease spreading and the size of disease outbreaks \cite{grassberger1983critical,bunde1985universality,herrmann1986geometrical,grenfell2001travelling,may2001infection,moreno2002epidemic,newman2002percolation,newman2002spread,lloyd2005superspreading,miller2007epidemic,kenah2007second,trapman2007analytical,davis2008abundance,lagorio2009effects,parshani2010epidemic,karrer2010message,neri2011effect,neri2011heterogeneity,ochab2011shift,zuzek2015epidemic,wang2017unification}. Nevertheless, models based on this approach also suffer from limitations: they typically either do not consider the dynamics of disease spreading and only treat the final stage of infection, or they also describe the dynamics of disease transmission using an \textit{ad hoc} macroscopic parameter that aggregates the influence of random and uncorrelated individual interactions. However, transmission is known to depend sensitively on the full history of infection, on specific individual behaviors \cite{stoneunexpected} including social distancing \cite{newman2002spread,lloyd2005superspreading,neri2011effect,peak2017comparing,scherer2020moving,maier2020effective}, and on interactions between different social networks \cite{zuzek2015epidemic}. Thus, the ability to accurately predict the temporal evolution of active infections in an overall population, as well as the full spatiotemporal features of disease spreading, remains limited. 

Here, we build on this previous work to develop an \textit{infection percolation} framework to model infectious disease spreading by applying $SIR$ dynamics to discrete interactions between members of a population. Our framework explicitly considers the full spatiotemporal history of infection as well as individual interactions in describing the spatial variation and individual dynamics of disease transmission. Simulations employing this framework show that disease spreads through a social network as a traveling wave of infection, followed by a traveling wave of recovery---consistent with previous predictions obtained using analytical $SIR$ modeling \cite{grenfell2001travelling,ruan2007spatial,lang2018analytic}. Analysis of these waves reveals general features of the total number of infections, maximal number of active infections, and the temporal evolution of active infections in a population, and clarifies how these features are determined by the interplay between disease transmission and recovery---consistent with the results of previous percolation simulations \cite{capala2017epidemics}. Our framework therefore provides a simple way to unify different findings previously obtained using disparate modeling approaches, and helps to clarify how disease spreading manifests for different diseases and containment strategies. Finally, as an example of how our framework can help to go beyond typical models of well-mixed populations that do not incorporate possible correlations in individual behavior, we demonstrate how disease spreading is strongly altered in a spatially heterogeneous social network.\\

\section*{Results}


\noindent \textbf{Development of the dynamic network model.} Our approach is inspired by dynamic network modeling of fluid-driven transport in heterogeneous media, which similarly seeks to predict spatiotemporal features of spreading in complex settings  \cite{stauffer2018introduction,masson2016fast,lu2019controlling,furuberg1988dynamics,roux1989temporal,pietronero1990percolation,moore2000epidemics,warren2001firewalls,sander2002percolation,sander2003epidemics,miller2009percolation,massaro2014epidemic}. We represent members of a population, all of whom are initially susceptible to infection, by sites of a network that are connected by bonds representing pairwise interactions between them (\textit{Methods}). We describe the intrinsic infectivity of a given disease throughout this network by the parameter $P^{*}$, where a high value of $P^{*}$ characterizes a highly infectious disease. Crucially, each interaction between individuals $ij$ is characterized by its own barrier to disease transmission $P_{\mathrm{th},ij}$ \cite{cardy1985epidemic}; this threshold explicitly describes the propensity of individual $i$, if infected, to transmit the disease to the susceptible individual $j$, and depends on individual behaviors such as social distancing between $i$ and $j$ \cite{neri2011effect,scherer2020moving,newman2002spread,maier2020effective}. Furthermore, if individuals $i$ and $j$ are infected and susceptible, respectively, the duration of disease transmission from $i$ to $j$ is given by $\Delta\tau_{ij}$; at the population scale, the mean of all the $\Delta\tau_{ij}$ can be thought of as the inverse of a macroscopic disease transmission rate. However, in what follows, we will describe our results in terms of the time durations, since our model utilizes the time duration, not rate, in calculations of different transmission events between individuals.

Previous studies indicate that disease is transmitted between individuals more rapidly, with shorter $\Delta\tau_{ij}$, for stronger infections $P^{*}$ and reduced barriers to transmission $P_{\mathrm{th},ij}$ \cite{crepey2006epidemic,wearing2005appropriate,van2013non,gomez1994estimation,degruttola1991modeling,aldrin2010stochastic,kristoffersen2009risk,stene2014transmission,tran2013highly,yun2015efficient,karlsson2015visualizing}. We therefore propose the \textit{ansatz} that disease is only transmitted if $P_{\mathrm{th},ij}<P^{*}$, with a transmission time $\Delta\tau_{ij}=\tau_{0}/f(P^*-P_{\mathrm{th},ij})$, where $f$ monotonically increases from 0 to 1 and $\tau_{0}$ is a characteristic minimal transmission time. This feature contrasts with previous percolation-based models, which often assume that the individual disease transmission times and barriers to disease transmission are either constant or drawn from two independent distributions. Instead, by explicit linking the individual  $\Delta\tau_{ij}$ and $P_{\mathrm{th},ij}$, our work provides a way to analyze the competing roles of individual disease infectivity and susceptibility in influencing population-scale disease spreading, as we show below. 

Finally, as in the classic $SIR$ model, we introduce the recovery duration $\tau_{r,i}$ after which individual $i$, if infected, transitions to a recovered state and can no longer infect other individuals. The mean of this quantity can be thought of as the inverse of a macroscopic infection recovery rate. However,  in what follows, we will describe our results in terms of the recovery duration---again, because our model utilizes the time duration, not rate, in calculations of different recovery events.

In nondimensional form, these control parameters---the infectivity, individual barriers to infection, disease transmission times, and recovery duration---are then represented by the parameters $\tilde{P}^{*}\equiv P^{*}/P_{\mathrm{th},\mathrm{max}}$, $\tilde{P}_{\mathrm{th},ij}\equiv P_{\mathrm{th},ij}/P_{\mathrm{th},\mathrm{max}}$, $\Delta\tilde{\tau}_{ij}\equiv\Delta\tau_{ij}/\tau_0=1/f(1-\tilde{P}_{\mathrm{th},ij}/\tilde{P}^*)$, and $\tilde{\tau}_{r,i}\equiv\tau_{r,i}/\tau_{0}$, respectively, where $P_{\mathrm{th},\mathrm{max}}$ is the maximal value of $P_{\mathrm{th},ij}$ in the population. Thus, this framework parses the role of a macroscopic transmission rate or reproduction number $R_0$ into appropriate parameters that are globally constant for a particular disease and parameters that vary by individual, providing a systematic way to incorporate measurable spreading parameters to model populations with spatial heterogeneities.

As a first step toward exploring the spatiotemporal features of disease spreading in this framework, we construct two-dimensional (2D) simulations implementing these deterministic rules. We represent the social network as a static square lattice with $N_{t}=10^{4}$ sites, though exploring more complex networks with small world and scale-free features \cite{watts1998collective,kleinberg2000navigation,barabasi2009scale,may2001infection} will be an important direction for future work. The disease is introduced at the central site at time $\tilde{\tau}\equiv\tau/\tau_{0}=0$. For simplicity, we take the barriers to disease transmission $\tilde{P}_{\mathrm{th},ij}$ to be undirected, with $\tilde{P}_{\mathrm{th},ij}=\tilde{P}_{\mathrm{th},ji}$; however, directed transmission is known to dramatically change disease spreading in some networks, and will be useful to explore in future implementations \cite{stoneunexpected}. The $\tilde{P}_{\mathrm{th},ij}$ have randomly assigned values that are chosen from a uniform distribution---and kept fixed for all computations performed for a given distribution---though we later test other distributions as well. We describe the individual dynamics of disease transmission using the simplest linear function,  $f=1-\tilde{P}_{\mathrm{th},ij}/\tilde{P}^*$, and take recovery to be an intrinsic property of a given disease, with $\tilde{\tau}_{r,i}$ set to a constant value $\tilde{\tau}_{r}$ throughout the lattice network. \\

\begin{figure*}
     \centering

    \includegraphics[width=\textwidth]{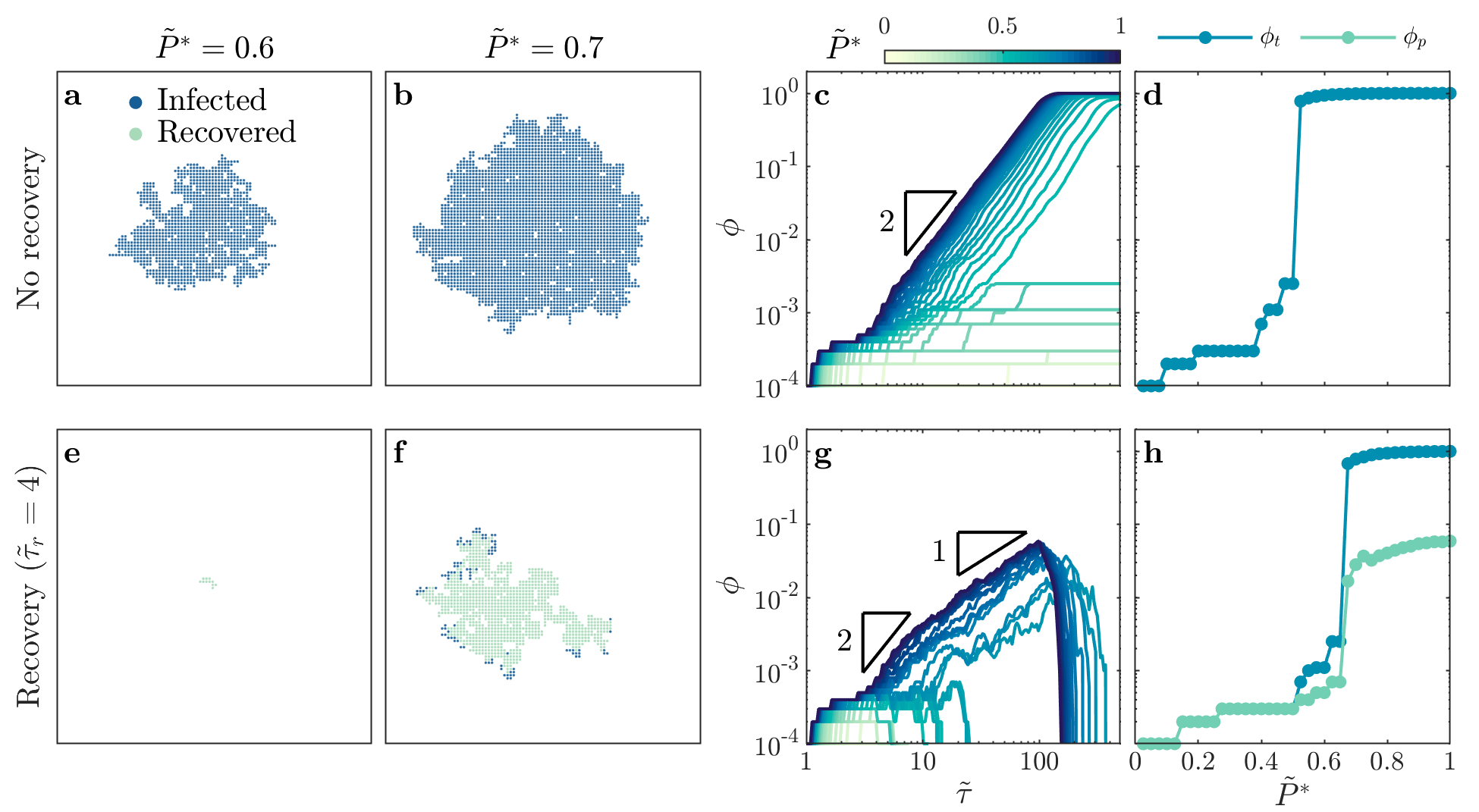}
\caption{\label{fig1} Recovery suppresses the onset and dynamics of disease spreading. For a population without recovery: \textbf{(a)} A disease with intermediate infectivity $\tilde{P}^{*}=0.6$ spreads in a spatially heterogeneous, ramified pattern, while \textbf{(b)} a disease with higher infectivity $\tilde{P}^{*}=0.7$ spreads in a more compact region. \textbf{(c)} Growth of infected fraction $\phi$ over time $\tilde{\tau}$ is hindered for diseases with low infectivity, but exhibits a generic quadratic scaling over time as indicated by the triangle, which eventually plateaus near unity for diseases with infectivity above a critical value. Infection growth is slower as $\tilde{P}^{*}$ decreases. \textbf{(d)} Total infected fraction $\phi_{t}$ exhibits an abrupt increase above a critical infectivity, $\tilde{P}_{c,0}\approx0.5$, indicating a percolation transition to an epidemic. For a population with recovery after $\tilde{\tau}_{r}=4$: \textbf{(e)} A disease with intermediate infectivity $\tilde{P}^{*}=0.6$ remains localized and does not spread, while \textbf{(f)} a disease with higher infectivity $\tilde{P}^{*}=0.7$ spreads in a spatially heterogeneous, ramified pattern. \textbf{(g)} Growth of infected fraction $\phi$ over time $\tilde{\tau}$ is hindered for diseases with low infectivity, but exhibits a generic quadratic scaling at intermediate times and linear scaling at longer times as indicated by the triangles, before reaching a peak value $\phi_{p}$ before dropping rapidly to zero for diseases with infectivity above a critical value. Infection growth is slower as $\tilde{P}^{*}$ decreases. \textbf{(h)} Total infected fraction $\phi_{t}$ and peak infected fraction $\phi_{p}$, which represent the total infected fraction at the end of a given simulation and the maximal infected fraction during the simulation, respectively, both exhibit an abrupt increase above a critical infectivity, $\tilde{P}_{c}\approx0.65$, that is larger than the recovery-free case---indicating that recovery suppresses the percolation transition to an epidemic. Images in \textbf{a-b}, \textbf{e-f} are for the same time $\tilde{\tau}=100$. All data are for a uniform distribution of individual interaction barriers $\tilde{P}_{\mathrm{th},ij}\in[0,1]$. Each curve in \textbf{c,g} and each point in \textbf{d,h} represents a separate simulation, all of which preserve the same lattice network structure of the $\tilde{P}_{\mathrm{th},ij}$ and only vary $\tilde{P}^{*}$ and $\tilde{\tau}_{r}$, enabling us to systematically test the influence of these two key parameters on disease spreading.}
\end{figure*}

\noindent\textbf{Infection percolation in a recovery-free population.} We first investigate the recovery-free case with $\tilde{\tau}_{r}\rightarrow\infty$. For a disease with low infectivity, $\tilde{P}^*\ll1$, only a small subset of individual interactions lead to transmission; thus, a large portion of the population becomes inaccessible to infection, and disease spreading is localized (Movie S1)---reminiscent of subcritical bond percolation. For sufficiently high infectivity, however, a sufficient number of individual interactions permit transmission for the disease to spread throughout the population---reminiscent of supercritical bond percolation. The example of $\tilde{P}^*=0.6$ (Movie S2) is shown at an intermediate time $\tilde{\tau}=100$ in Figure \ref{fig1}a. For this infectivity, the disease spreads in a spatially heterogeneous, ramified pattern, reminiscent of supercritical bond percolation near the percolation threshold. This heterogeneous spreading leads to the formation of discrete clusters of bypassed individuals who remain uninfected (white in Fig. \ref{fig1}a). For diseases of higher infectivity, disease spreading is more compact: the leading front of the infected population becomes more smooth, resulting in fewer and smaller uninfected clusters, as shown for the example of $\tilde{P}^*=0.7$ in Fig. \ref{fig1}b and Movie S3. Clearly, disease infectivity strongly influences the spatiotemporal features of spreading.

To gain further insight into disease spreading, we repeat these simulations for a broad range of $\tilde{P}^*$, and characterize the infection growth dynamics by measuring the time-dependent infected fraction of the population, $\phi$. We again observe a bifurcation in infection behavior. For low infectivity, the disease spreads slowly (light curves in Fig. \ref{fig1}c), ultimately only infecting a total fraction $\phi_{t}\ll1$ (points with $\tilde{P}^*<0.5$ in Fig. \ref{fig1}d). By contrast, for sufficiently high infectivity, the disease spreads rapidly with $\phi\sim\tilde{\tau}^{2}$ (dark curves in Fig. \ref{fig1}c), ultimately infecting nearly the entire population (points for $\tilde{P}^*>0.5$ in Fig. \ref{fig1}d). The abrupt onset of rapid spreading throughout the population at a critical infectivity $\tilde{P}_{c,0}\approx0.5$ again suggests that disease spreading can be described as a dynamic process of percolation through infected bonds, consistent with previous calculations \cite{bunde1985universality,herrmann1986geometrical,may2001infection,newman2002percolation,newman2002spread,moreno2002epidemic,lloyd2005superspreading,miller2007epidemic,kenah2007second,trapman2007analytical,davis2008abundance,lagorio2009effects,karrer2010message,parshani2010epidemic,neri2011heterogeneity,neri2011effect,ochab2011shift,zuzek2015epidemic,wang2017unification}. This suggestion is further confirmed by the value of $\tilde{P}_{c,0}$, which coincides with the critical probability of bond percolation on the 2D square lattice \cite{kesten1980critical,mccarthy1987invasion,ochab2011shift}. We find no discernible difference in these results when the simulations are performed with lattice networks that are up to two orders of magnitude larger (Fig. S1), suggesting that our results are not strongly sensitive to finite-size effects.

\begin{SCfigure*}[\sidecaptionrelwidth][t]
    \sidecaptionvpos{figure}{t}
    \centering
    \includegraphics[width=0.7\textwidth]{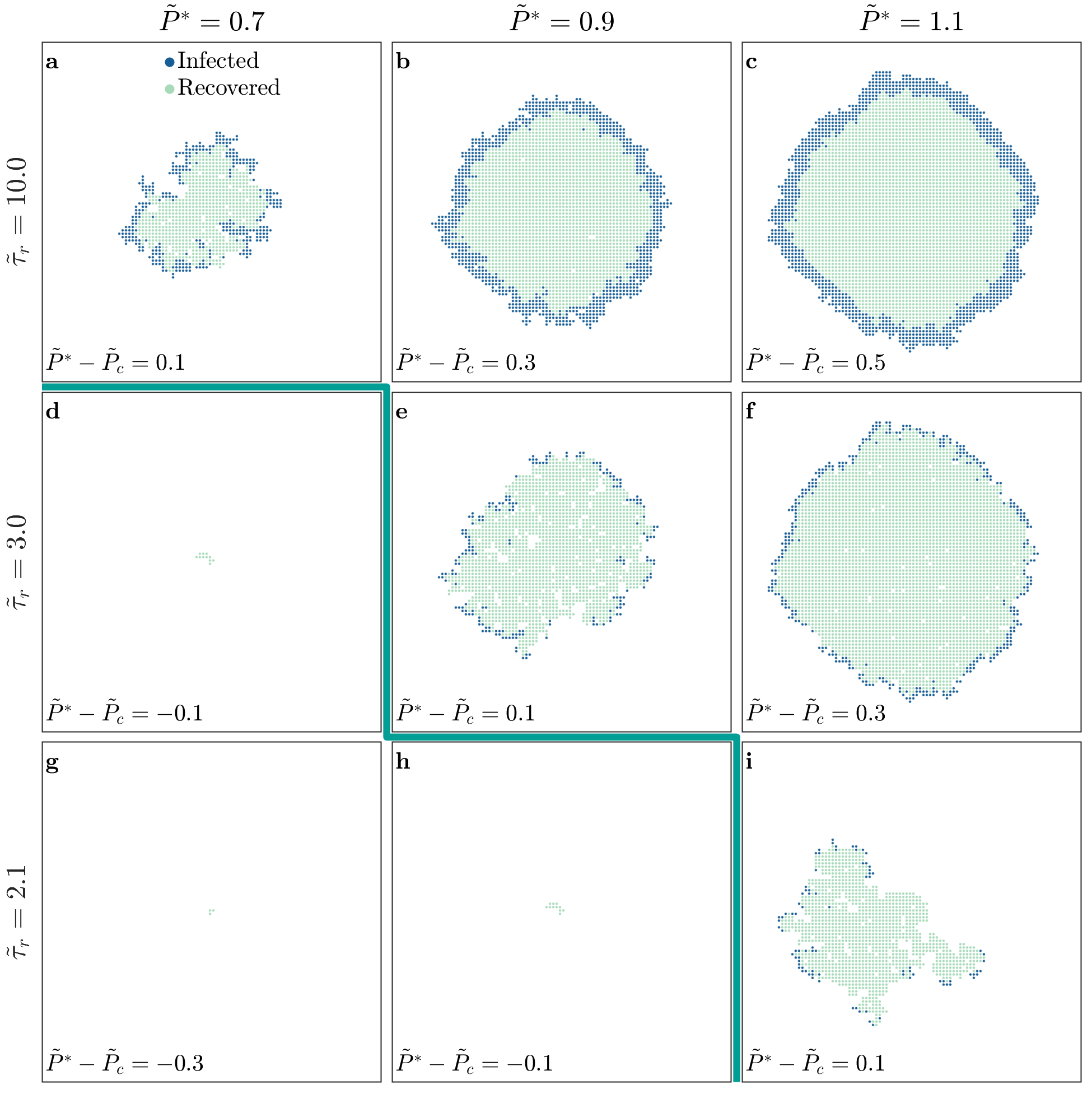}
    \caption{\label{fig2} The competition between infection and recovery determines the onset and pattern of disease spreading. For a population with a longer recovery duration $\tilde{\tau}_r=10.0$, \textbf{(a)} a disease with lower infectivity $\tilde{P}^{*}=0.7$ spreads in a ramified pattern, while diseases with higher infectivities $\tilde{P}^{*}=0.9$ and $1.1$ \textbf{(b-c)} spread in a more compact manner. The leading front of the infected population is trailed by an inner region of recovery, leading to the formation of a circular traveling pulse of infection. For a shorter recovery duration of $\tilde{\tau}_r=3.0$, \textbf{(d-e)} the threshold infectivity for appreciable disease spreading is larger, and \textbf{(e-f)} the pulse of infection is thinner. These effects are even more pronounced for the shortest recovery duration of $\tilde{\tau}_r=2.1$, as shown in \textbf{(g-i)}. The thick green line indicates the transition to infection percolation. All images are shown for the same time $\tilde{\tau}=75$ and all data are for a uniform distribution of individual interaction barriers $\tilde{P}_{\mathrm{th},ij}\in[0,1]$ preserving the same lattice network structure of the $\tilde{P}_{\mathrm{th},ij}$ throughout, thus demonstrating how varying $\tilde{P}^{*}$ and $\tilde{\tau}_{r}$ influences disease spreading.}
\end{SCfigure*}

Why does infection growth show quadratic scaling in time for $\tilde{P}^{*}>\tilde{P}_{c,0}$? At the leading front of the infected population, new individuals are infected over a range of transmission times $\Delta\tilde{\tau}_{ij}=1/(1-\tilde{P}_{\mathrm{th},ij}/\tilde{P}^*)\in[1,\infty)$, leading to heterogeneous disease spreading. However, as $\tilde{P}^{*}$ increases, a greater proportion of disease transmission between individuals occurs in the shortest possible time  $\tau_{0}$, corresponding to $\Delta\tilde{\tau}_{ij}=1$, resulting in more compact spreading (Fig. \ref{fig1}b). Hence, the leading front of the infected population spreads radially outward on the 2D social network lattice at a maximal rate of 1 new individual per $\tau_{0}$, therefore spanning an overall region with a maximal radius of $\tilde{\tau}$ individuals. The maximal infected fraction of the population is then given directly by the area of this infected region: $\phi\approx\pi\tilde{\tau}^{2}/N_{t}$, consistent with a previous result in percolation theory \cite{burlatsky1985growth} and yielding the quadratic scaling shown in Fig. \ref{fig1}c. As this infected region spreads, it eventually reaches the boundary and spans the entire population at a shortest possible time of $\tilde{\tau}_{f,0}\approx\sqrt{N_{t}/2}=71$ (\textit{SI text})---in good agreement with the onset of the plateau in $\phi$ at $\tilde{\tau}\approx100$ for the highest $\tilde{P}^{*}$ shown in Fig. \ref{fig1}c. As $\tilde{P}^{*}$ decreases, we expect that the infected region spreads at a slower rate, as a greater proportion of disease transmission occurs at $\Delta\tilde{\tau}_{ij}>1$---also in good agreement with the variation of the curves in Fig. \ref{fig1}c. The variability in these curves reflects the increasing importance of the variability in the individual barriers; different simulations employing different choices of the randomly-chosen $\tilde{P}_{\mathrm{th},ij}$ would yield slightly different dynamics. However, in all cases, we expect that decreasing $\tilde{P}^{*}$ leads to slower transmission as a greater proportion of disease transmission occurs at $\Delta\tilde{\tau}_{ij}>1$. \\


\noindent\textbf{Infection percolation in a population with recovery.} How do these results change when infected individuals can recover? To address this question, we perform the same simulations as in Fig. \ref{fig1}a--d, but with $\tilde{\tau}_{r}=4$. This modification markedly alters disease spreading. For $\tilde{P}^*=0.6$, the spread of disease is quickly quenched by recovery \cite{karrer2010message}, and only a few individuals are ever infected (Fig. \ref{fig1}e, Movie S4). Thus, even above the critical infectivity for percolation in the recovery-free case, $\tilde{P}_{c,0}\approx0.5$, recovery gives rise to subcritical spreading behavior. For a higher infectivity of $\tilde{P}^*=0.7$, the disease does continue to spread, eventually reaching the boundaries of the population as in supercritical bond percolation. However, unlike the recovery-free case, it does so in a spatially heterogeneous, ramified pattern (Fig. \ref{fig1}f, Movie S5). Close inspection of the spatiotemporal pattern of infection and recovery reveals the underlying cause: recovery of infected individuals before they can transmit the disease shields clusters of individuals who would have otherwise been infected, as exemplified by the large uninfected region in the top right of Fig. \ref{fig1}f, which was heavily infected in the recovery-free case shown in Fig. \ref{fig1}b. Together, these results hint that recovery suppresses infection percolation. 
The competition between infection and recovery also drastically alters the time evolution of the infected fraction $\phi$. For low infectivity, recovery is sufficiently fast to quench the spread of disease (light curves in Fig. \ref{fig1}g). As a result, the total fraction of the population ever infected $\phi_{t}\ll1$ (points with $\tilde{P}^*<0.65$ in Fig. \ref{fig1}h). By contrast, for sufficiently high infectivity, the disease initially spreads rapidly, first with $\phi\sim\tilde{\tau}^{2}$ as in the recovery-free case and then with $\phi\sim\tilde{\tau}^{1}$ (dark curves in Fig. \ref{fig1}g). As time progresses  $\phi$ eventually reaches a peak value $\phi_{p}$ before dropping rapidly to zero as the entire population recovers. Both $\phi_{t}$ and $\phi_{p}$ increase precipitously above the critical infectivity $\tilde{P}_{c}\approx0.65>\tilde{P}_{c,0}$ (Fig. \ref{fig1}h), again indicating that recovery suppresses infection percolation. Again, the variability in the curves in Fig. 1g reflects the variability in the individual barriers; different simulations employing different choices of the randomly-chosen $\tilde{P}_{\mathrm{th},ij}$ would yield slightly different dynamics. However, in all cases, we expect that decreasing $\tilde{P}^{*}$ leads to slower transmission as a greater proportion of disease transmission occurs at $\Delta\tilde{\tau}_{ij}>1$.


To further explore the competition between infection and recovery, we inspect the spatiotemporal patterns of both using simulations performed at several different values of $\tilde{P}^{*}$ and $\tilde{\tau}_{r}$, with snapshots all taken at $\tilde{\tau}=75$ shown in Fig. \ref{fig2}. For a large value of $\tilde{\tau}_{r}=10$, the critical infectivity for percolation is $\tilde{P}_{c}\approx0.6$; for $\tilde{P}^{*}$ slightly above this value, disease spreading is heterogeneous, with recovered individuals again shielding clusters of individuals who would have otherwise been infected (Fig. \ref{fig2}a). For even higher $\tilde{P}^{*}$, disease spreading becomes more compact (Figs. \ref{fig2}b-c). Moreover, the leading front of the infected population is trailed by an inner compact region of recovery, leading to the formation of a wide circular pulse of infection that travels outward through the population (dark blue region in Figs. \ref{fig2}b-c). This feature is notably similar to the traveling pulses of infection predicted by extensions of the classic $SIR$ model \cite{grenfell2001travelling,ruan2007spatial,lang2018analytic}---highlighting the ability of our framework to reproduce previously-reported continuum-scale phenomena. We observe similar behavior at smaller values of $\tilde{\tau}_{r}$, but shifted to increased values of $\tilde{P}^{*}$ (Figs. \ref{fig2}d-i): recovery can increasingly quench the spread of disease as it becomes faster relative to infection transmission. As a result, the critical infectivity for percolation, $\tilde{P}_{c}$, shifts to higher values (thick green line in Fig. \ref{fig2}), further confirming that recovery suppresses infection percolation. Recovery also strongly impacts the number of infected individuals: as $\tilde{\tau}_{r}$ decreases, the thickness of the pulse of infected individuals decreases (compare Figs. \ref{fig2}c, f, i).

\begin{figure*}
\centering
\includegraphics[width=\textwidth]{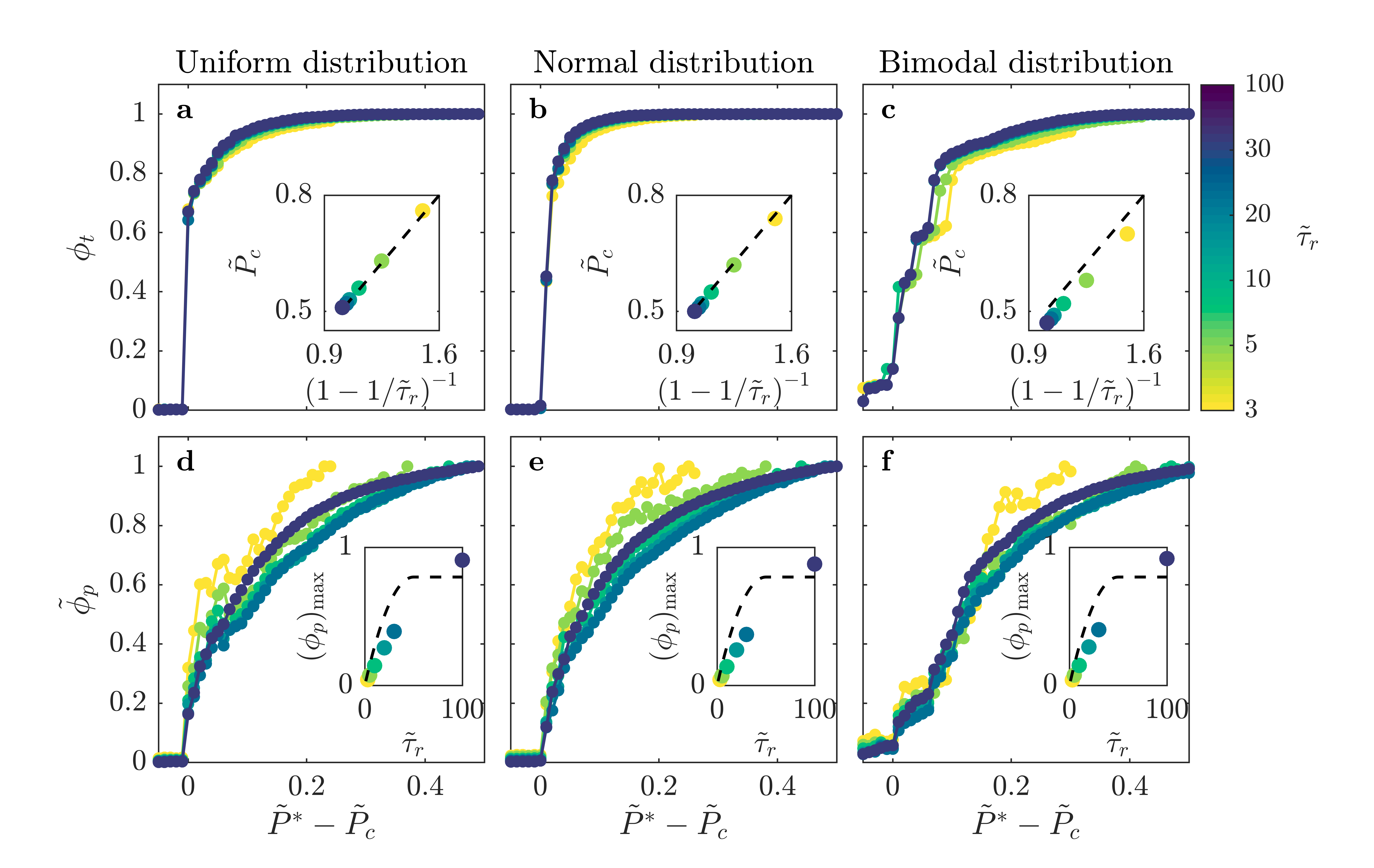}
\caption{\label{fig3} Amount of infection shows a general dependence on disease infectivity across populations with different recovery durations and distributions of individual interaction barriers. \textbf{(a-c)} Data for the total infected fraction $\phi_{t}$ align when plotted as a function of the shifted disease infectivity $\tilde{P}^{*}-\tilde{P}_{c}$; insets show that the critical infectivity $\tilde{P}_{c}$ is given by the relation in Eq. \ref{eq1} (dashed lines). \textbf{(d-f)} Data for the rescaled peak infected fraction $\tilde{\phi}_{p}\equiv\phi_{p}/(\phi_p)_\mathrm{max}$ show reasonable alignment when plotted as a function of the shifted disease infectivity $\tilde{P}^{*}-\tilde{P}_{c}$; insets show that the theoretical $\phi(\tilde{\tau}_{p})$ calculated using Eq. \ref{eq2} (dashed lines) provides a reasonable approximation to the maximal peak fraction $(\phi_p)_\mathrm{max}$ determined for $\tilde{P}^{*}=2$. Intriguingly, all the datasets appear to approximately converge as $\tilde{\tau}_{r}$ increases. All three distributions span $\tilde{P}_{\mathrm{th},ij}\in[0,1]$; the normal distribution is centered at 0.5 and has standard deviation = 0.25, while the bimodal is constructed from two normal distributions centered at 0 and 1, both with standard deviation = 0.25. Each point represents a separate simulation, all of which preserve the same lattice network structure of the $\tilde{P}_{\mathrm{th},ij}$ for a given distribution and only vary $\tilde{P}^{*}$ and $\tilde{\tau}_{r}$, enabling us to systematically test the influence of these two key parameters on disease spreading.}
\end{figure*}

These results demonstrate the key influence of recovery on disease spreading\textcolor{black}{; we therefore examine the underlying model further to develop and test analytical relations that confirm the internal consistency of our simulations.} We first focus on the observation that faster recovery can quench the spread of disease. At the leading front of the infected population, new individuals are infected only when the infection transmission time $\Delta\tilde{\tau}_{ij}=1/(1-\tilde{P}_{\mathrm{th},ij}/\tilde{P}^{*})$ is shorter than the recovery time $\tilde{\tau}_{r}$; otherwise, an infected individual recovers from the disease before they are able to transmit it to a neighbor. Thus, only individual interactions with $\tilde{P}_{\mathrm{th},ij}<\tilde{P}^{*}(1-\tilde{\tau}_{r}^{-1})$ can transmit disease. For a disease to continually spread throughout the population, a sufficient number of these interactions must permit disease transmission, as given by the critical probability of bond percolation $\tilde{P}_{c,0}$; that is, infection percolates only when $\tilde{P}_{c,0}<\tilde{P}^{*}(1-\tilde{\tau}_{r}^{-1})$. The critical infectivity in a population with recovery is then directly given by the relation
\begin{equation}
    \tilde{P}_{c}=\frac{\tilde{P}_{c,0}}{1-\tilde{\tau}_{r}^{-1}} ,
    \label{eq1}
\end{equation}
\noindent which increases with $\tilde{\tau}_{r}$: faster recovery suppresses infection percolation. This \textcolor{black}{dependence} is in good agreement with the shift in the critical infectivity shown in Fig. \ref{fig1}d, h and by the thick green line in Fig. \ref{fig2}.

To further \textcolor{black}{confirm the internal consistency of our simulations,} we \textcolor{black}{explore} values of $\tilde{\tau}_r$ spanning nearly two orders of magnitude. For each simulation, we vary the disease infectivity $\tilde{P}^{*}$ and identify the critical infectivity $\tilde{P}_{c}$ at which the total fraction of the population ever infected $\phi_{t}$ abruptly increases, similar to the curves shown in Fig. \ref{fig1}d, h. Consistent with our expectation, all the data for different $\tilde{\tau}_{r}$ show similar growth when plotted as a function of the shifted $\tilde{P}^{*}-\tilde{P}_{c}$ (Fig. \ref{fig3}a). Moreover, the variation of the critical infectivity $\tilde{P}_{c}$ with $\tilde{\tau}_{r}$ shows excellent agreement with Eq. \ref{eq1} (dashed line, Fig. \ref{fig3}a inset). 

Finally, we test the generality of this \textcolor{black}{relation} by exploring two other distributions of $\tilde{P}_{\mathrm{th},ij}$: a normal distribution, representing a population with random and uncorrelated interactions whose variation is distributed about a single mean value, and a bimodal distribution, representing a population with distinct high-risk and low-risk subpopulations arising from, for example, differences in compliance with public health interventions. Our central findings that the onset and dynamics of disease spreading are regulated by the interplay between disease transmission and recovery are consistent across the different distributions tested. Specifically, for all the distributions tested, we observe a similar abrupt increase in $\phi_{t}$ with $\tilde{P}^{*}$ above a critical infectivity $\tilde{P}_{c}$, with excellent alignment of all the data for different $\tilde{\tau}_{r}$ (Figs. \ref{fig3}a-c). In all cases, the value of $\tilde{P}_{c}$ determined shows close agreement with Eq. \ref{eq1} (Figs. \ref{fig3}a-c insets). A bimodal population in which half of the contacts between members maintain high barriers to transmission while the other interspersed connections do not---such as through strong or weak social distancing, respectively---has $\tilde{P}_{c}$ slightly less than this \textcolor{black}{relation}, indicating that it is slightly more susceptible to infection. \\

\begin{figure*}
\includegraphics[width=6in]{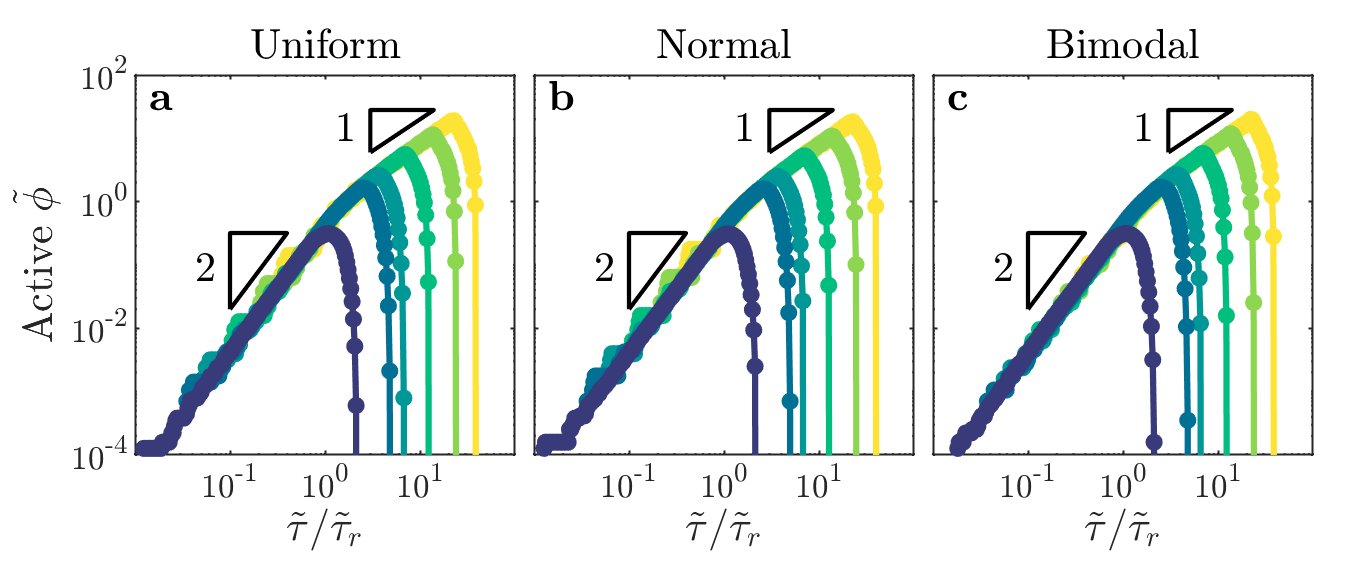}
\caption{\label{fig4} Total fraction of active infections in a population shows general dynamics across populations with different recovery durations and distributions of individual interaction barriers. \textbf{(a-c)} Growth of rescaled infected fraction $\tilde{\phi}\equiv\phi(\tilde{\tau})/\phi(\tilde{\tau}_{r})$ with rescaled time $\tilde{\tau}/\tilde{\tau}_{r}$ shows general quadratic to linear scaling as indicated by the triangles, followed by a drop to zero at $\tilde{\tau}_{f}/\tilde{\tau}_{r}\approx\sqrt{N_{t}/2}/\tilde{\tau}_{r}+1$. Data shown are for the largest disease infectivity tested, $\tilde{P}^{*}=2$. Distributions of individual interaction barriers and colors indicating different recovery durations $\tilde{\tau}_{r}$ are the same as in Fig. \ref{fig3}. Each point represents a separate simulation, all of which preserve the same lattice network structure of the $\tilde{P}_{\mathrm{th},ij}$ for a given distribution and only vary $\tilde{\tau}_{r}$.}
\end{figure*}

\newpage\noindent\textbf{General scaling of infection growth dynamics.} We next focus on the observation that, for high $\tilde{P}^{*}$, the leading front of the infected population is trailed by an inner region of recovery. At the leading front of the infected population, the shortest possible disease transmission time is again $\Delta\tilde{\tau}_{ij}=1$; therefore, the leading front spans an overall region with a maximal radius of $\tilde{\tau}$ individuals, as in the recovery-free case. For short times $\tilde{\tau}<\tilde{\tau}_{r}$, these individuals have not yet recovered; hence, we expect that $\phi\approx\pi\tilde{\tau}^{2}/N_{t}$ again in this regime, in agreement with the short-time quadratic scaling shown in Fig. \ref{fig1}g as well as the recovery-free quadratic scaling shown in Fig. \ref{fig1}c. At longer times $\tilde{\tau}\geq\tilde{\tau}_{r}$, infected individuals begin to recover, forming an inner region of recovery. The leading front of this region spreads at the same rate as the leading front of the infected population; however, it only spans a maximal radius of $\tilde{\tau}-\tilde{\tau}_{r}$ individuals. The maximal infected fraction of the population is then given by the area between the leading fronts of the infected and recovered populations: $\phi\approx\pi[\tilde{\tau}^{2}-(\tilde{\tau}-\tilde{\tau}_{r})^{2}]/N_{t}=\pi\tilde{\tau}_{r}^{2}(2\tilde{\tau}/\tilde{\tau}_{r}-1)/N_{t}$, yielding the measured intermediate-time linear scaling (Fig. \ref{fig1}g). Furthermore, as the infected region spreads, it eventually reaches  the boundary, followed by the growing region of recovery $\tilde{\tau}_{r}$ later. We therefore expect that the infected fraction $\phi$ will peak at a time $\tilde{\tau}_{p}\approx\sqrt{N_{t}}/2$ before dropping rapidly and reaching zero at $\tilde{\tau}_{f}\approx\sqrt{N_{t}/2}+\tilde{\tau}_{r}$, corresponding to $\tilde{\tau}_{p}\approx50$ and $\tilde{\tau}_{f}\approx75$, respectively, for $\tilde{\tau}_{r}=4$ (\textit{SI text})---in good agreement with the results shown in Fig. \ref{fig1}g. Together, these calculations yield a general \textcolor{black}{expression} for the full time evolution of the infected fraction of a population in the limit of high $\tilde{P}^{*}$ well above the threshold $P_c$, in excellent agreement with the data shown in Fig. \ref{fig1}g:\begin{equation}
\phi(\tilde{\tau}) \approx 
  \begin{cases}
    \pi\tilde{\tau}^{2}/N_{t} & \text{when} ~\tilde{\tau}<\mathrm{min}(\tilde{\tau}_{r},\tilde{\tau}_{p})\\
   \left[\pi\tilde{\tau}_{r}^{2}(2\tilde{\tau}/\tilde{\tau}_{r}-1)\right]/N_t & \text{when} ~\tilde{\tau}_{r}\leq\tilde{\tau}\leq\tilde{\tau}_{p}
  \end{cases}
\label{eq2}
\end{equation}
As $\tilde{P}^{*}$ decreases, we again expect that the infected region spreads at a slower rate, prolonging $\tilde{\tau}_{p}$---also in good agreement with the different curves shown in Fig. \ref{fig1}g. 

Eq. \ref{eq2} also provides an approximation of the peak infected fraction, $\phi_{p}=\phi(\tilde{\tau}_{p})$, in the limit of high $\tilde{P}^{*}$; for the case of $\tilde{\tau}_{r}=4$, we estimate $\phi_{p}\approx0.1$, consistent with the data shown in Fig. \ref{fig1}g. To further \textcolor{black}{examine} this \textcolor{black}{relation}, we analyze the results of all the simulations described in Fig. \ref{fig3} with varying $\tilde{\tau}_{r}$. For each simulation, we determine the maximal value of $\phi_{p}$, $(\phi_{p})_\mathrm{max}$, at $\tilde{P}^{*}=2$, the highest value tested. Consistent with our expectation, $\phi_{p}=\phi(\tilde{\tau}_{p})$  calculated using Eq. \ref{eq2} (dashed line, Fig. \ref{fig3}d inset), provides a reasonable approximation to the measured $(\phi_{p})_\mathrm{max}$ (data points, Fig. \ref{fig3}d inset). Furthermore, all the data for $\tilde{\phi}_{p}\equiv\phi_{p}/(\phi_{p})_\mathrm{max}$ show reasonable alignment when plotted as a function of the shifted $\tilde{P}^{*}-\tilde{P}_{c}$ (Fig. \ref{fig3}d), with some deviation for the smallest $\tilde{\tau}_{r}$ likely arising from geometric effects not taken into account in our simple estimate of $\phi_{p}=\phi(\tilde{\tau}_{p})$. We again test the generality of these results by exploring their applicability to a normal distribution and a bimodal distribution of $\tilde{P}_{\mathrm{th},ij}$; the results are closely similar in all three cases (Figs. \ref{fig3}d-f). Taken together, all of the results shown in Figs. \ref{fig1}-\ref{fig3} support the validity of our scaling \textcolor{black}{relations}.

\begin{figure*}
\centering
\includegraphics[width=6.8in]{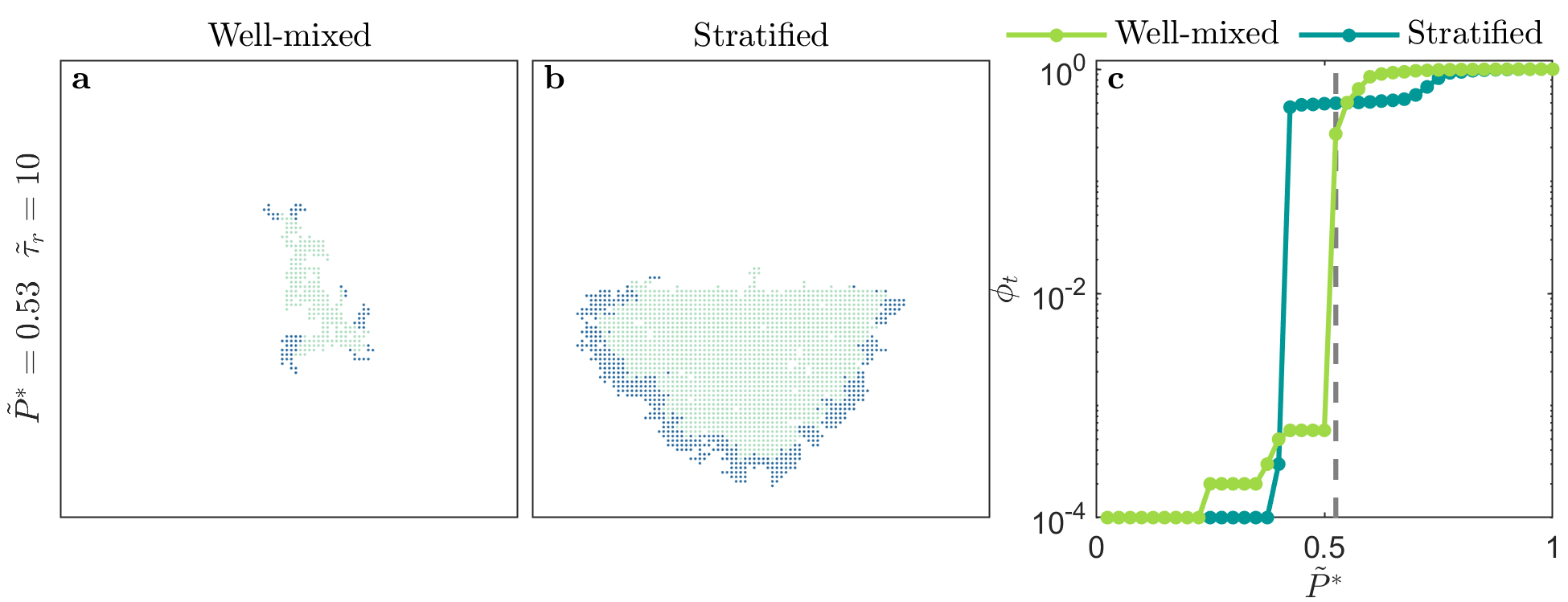}
\caption{\label{SI Fig2} Spatial heterogeneity strongly alters disease spreading. \textbf{(a)} Homogeneously-mixed population i.e. the distribution of $\tilde{P}_{\mathrm{th},ij}\in[0,1]$ is bimodal, constructed from two normal distributions centered at 0 and 1, both with standard deviation = 0.25, and with $\tilde{P}_{\mathrm{th},ij}$ values randomly distributed through the lattice network. In this case, the disease slowly spreads through the population in a spatially heterogeneous manner. \textbf{(b)} Stratified population with the exact same overall distribution of $\tilde{P}_{\mathrm{th},ij}\in[0,1]$, but with the individuals comprising the normal distributions centered at 0 (more susceptible) and 1 (less susceptible) isolated to the bottom and top halves of the lattice network, respectively. In this case, disease spreading is dramatically worsened; the total amount of infection is larger than the homogeneously-mixed case. Moreover, disease spreading is stronger in the more susceptible subpopulation and is weaker in the less susceptible subpopulation. Images show the case of a disease with infectivity $\tilde{P}^{*}=0.525$, a population with recovery duration $\tilde{\tau}_{r}$=10, and time $\tilde{\tau}=100$. \textbf{(c)} Variation of the total infected fraction $\phi_t$ with the disease infectivity $\tilde{P}^{*}$ for both populations. For diseases of intermediate infectivity, $\tilde{P}^*=0.425$ to $0.55$, disease spreading is stronger in the stratified population. For diseases of higher infectivity, $\tilde{P}^*=0.55$ to $0.85$, disease spreading is stronger in the homogeneously-mixed population. Each point represents a separate simulation, all of which preserve the same lattice network structure of the $\tilde{P}_{\mathrm{th},ij}$ for either the homogeneously-mixed or stratified case and only vary $\tilde{P}^{*}$. Dashed line indicates $\tilde{P}^*=0.525$, corresponding to the images in \textbf{(a-b)}. 
}
\end{figure*}

As a final \textcolor{black}{exploration} of \textcolor{black}{these relations}, we use all of our simulations to directly test the \textcolor{black}{general expression} for the infected fraction $\phi$ given by Eq. \ref{eq2}. All of the data collapse onto this scaling curve for all values of $\tilde{\tau}_{r}$ and distributions of $\tilde{P}_{\mathrm{th},ij}$ tested (Figs. \ref{fig4}a-c). Furthermore, the peak in $\phi$ and its eventual drop to zero occur at values of $\tilde{\tau}/\tilde{\tau}_{r}$ close to the predicted values $\tilde{\tau}_{p}/\tilde{\tau}_{r}\approx\sqrt{N_{t}}/2\tilde{\tau}_{r}$ and $\tilde{\tau}_{f}/\tilde{\tau}_{r}\approx\sqrt{N_{t}/2}/\tilde{\tau}_{r}+1$, respectively. Thus, the close agreement between all the data and the theoretical prediction confirm the \textcolor{black}{internal consistency of our simulations with the scaling relations}. We note that \textcolor{black}{the analytical expression} represents the solution for the high infectivity limit with high $\tilde{P}^{*}$; however, as indicated by the data in Figs. \ref{fig1}g and \ref{fig4}a-c, it provides a good approximation for the spreading dynamics of a broad range of diseases with $\tilde{P}^{*}\gtrsim P_{c}$. \textcolor{black}{Intriguingly, similar quadratic power-law scalings have been reported for the initial regional epidemic spreads of COVID-19 \cite{maier2020effective}, though explaining the full dynamics of this pandemic involves additional complexities that we do not consider here.}

\noindent\textbf{Extension to spatially heterogeneous networks.} While our simple implementation thus far considered distributions of $\tilde{P}_{\mathrm{th},ij}$ that were homogeneously mixed over the social network lattice, our framework allows for spatially heterogeneous networks to be constructed to model disease spreading in specific community structures \textcolor{black}{\cite{glass2004effect,van2001measles,guimera2003self}}. As an example of this idea, we test disease spreading from the boundary between two stratified subpopulations, each on a separate lattice with a different mean $\tilde{P}_{\mathrm{th},ij}$, representing differences in susceptibility to infection that can arise from differences in containment strategies or in socioeconomic factors \cite{grassly2008mathematical,glass2004effect,van2001measles,ball2006optimal}. Depending on the viral infectivity $\tilde{P}^{*}$, we find that stratification can either markedly exacerbate or hinder the overall amount of infection, and the total amount of infection in each subpopulation, compared to the homogeneously-mixed case (Fig. \ref{SI Fig2}). Specifically, for diseases of intermediate infectivity, disease spreading is stronger in the stratified population---the total infected fraction $\phi_{t}$ is two orders of magnitude larger than in the homogeneously-mixed case---due to the earlier onset of infection percolation in the more susceptible subpopulation. However, for diseases of higher infectivity, disease spreading is slightly stronger in the homogeneously-mixed population; in the stratified case, the less susceptible subpopulation buffers against continued spread of disease. These results exemplify how our framework provides a way to directly assess the critical role played by community structure in the spread of disease. More extreme forms of heterogeneity---for example with small-world connections between subpopulations, or power law distributions in susceptibility---can easily be incorporated within this framework, and will be an interesting direction for future research. For example, as suggested in previous work \cite{lang2018analytic}, we anticipate that long-range connections between regions with disparate susceptibilities will increasingly fragment traveling waves of infection.


\section*{Discussion}
The framework presented here provides a direct way to merge the temporal dynamics underlying the classic $SIR$ model with a network representation of all the discrete interactions between members of a population. We demonstrate this principle using 2D lattice simulations implementing a simplified form of this framework. Our simulations reveal that, for diverse diseases and populations, disease spreading can be understood as a process of infection percolation through a social network. By testing different recovery durations and distributions of individual interaction parameters, we find that the onset and dynamics of spreading are determined by the interplay between disease transmission and recovery at the scale of individual interactions. This finding thus builds on the rich body of previous work exploring disease spreading through the lens of percolation theory \cite{grassberger1983critical,bunde1985universality,herrmann1986geometrical,grenfell2001travelling,may2001infection,moreno2002epidemic,newman2002percolation,newman2002spread,lloyd2005superspreading,miller2007epidemic,kenah2007second,trapman2007analytical,davis2008abundance,lagorio2009effects,parshani2010epidemic,karrer2010message,neri2011effect,neri2011heterogeneity,ochab2011shift,wang2017unification}. 

Guided by these insights, we develop a scaling theory that yields general predictions for the total number of infections, maximal number of active infections, and the temporal evolution of active infections in a population. Importantly, our scaling theory clarifies how these predictions can be applied to different diseases, with varying infectivities $\tilde{P}^{*}$ and recovery durations $\tilde{\tau}_{r}$, and different populations, with varying distributions of the individual barriers to interaction $\tilde{P}_{\mathrm{th},ij}$ \textit{e.g.} due to different implementations of containment strategies like social distancing. 

Our simulations represent a first step toward implementing this framework. Thus, we have necessarily made a number of simplifying assumptions that can be relaxed in future extensions of this work. For example, our simulations consider a social network represented by a static, 2D, square lattice---while in reality, social networks are dynamic and have more complex structures. While previous work suggests that the assumption of a static network may not greatly alter disease spreading \cite{ochab2011shift}, our framework can be implemented on dynamically-changing networks using a time-dependent matrix of individual interactions $(\tilde{P}_{\mathrm{th},ij})$. Furthermore, while previous work suggests that disease spreading is well-approximated by spreading on a 2D network covering the Earth's surface \cite{newman2002percolation,bailey2000saprotrophic,otten2004empirical}, networks of different dimensionality $d$ can also be implemented---which would result in a more general form of Eq. \ref{eq2} with $\phi\sim\tilde{\tau}^{d}$ at short times and $\phi\sim\tilde{\tau}^{d-1}$ at longer times. Previous work also suggests that social networks can have broadly-distributed degree distributions \cite{barabasi2009scale,may2001infection}, small world connections \cite{watts1998collective,kleinberg2000navigation}, or multiple layers \cite{zuzek2015epidemic} unlike the fixed 2D connectivity of our lattice; \textcolor{black}{moreover, disease transmission can occur among groups, not just pairs of individuals, which would require consideration of a network with higher-order interactions \cite{arenasabrupt}}. Incorporating these features into our framework will be a valuable direction for future work. Indeed, the random long-range connections that arise in small-world networks could disrupt the traveling waves of infection and recovery we observed in our simple network, or even seed new traveling waves of infection and recovery, potentially leading to richer dynamics. Our treatment of individual interactions can also be extended in future work. For example, the barriers to disease transmission $\tilde{P}_{\mathrm{th},ij}$ need not be undirected, and the recovery durations $\tilde{\tau}_{r,i}$ need not be constant throughout the network. Furthermore, our representation of the disease transmission function $f=1-\tilde{P}_{\mathrm{th},ij}/\tilde{P}^{*}$ represents the simplest linear function, and can instead be replaced by a more complex form that incorporates the sophisticated dynamics of transmission specific to different diseases \cite{woolhouse2005epidemiological}.

Our framework could also enable straightforward assessment of the efficacy of different public health policies. For example, the implementation of strong social distancing results in an increase in $P_{\mathrm{th},\mathrm{max}}$, leading to a reduction in $\tilde{P}^{*}$ and hence a reduction in the peak infected fraction of the population, $\phi_{p}$ (Figs. \ref{fig3}d-f)---consistent with previous work \cite{maier2020effective,ball2006optimal,valdez2013social}. Alternatively, the development of better treatments shortening the recovery duration $\tilde{\tau_{r}}$ also leads to a reduction in the maximal peak infected fraction of the population, $(\phi_{p})_{\mathrm{max}}$ (Figs. \ref{fig3}d-f, insets); it also hastens the transition to slow linear scaling and eventual decline of infection growth (Figs. \ref{fig4}a-c)---again consistent with previous work \cite{peak2017comparing,chen2020convalescent}. The influence of other factors that are documented to impact disease spreading---e.g. seasonality of infectivity \textcolor{black}{\cite{viboud2005multinational,koelle2005pathogen,kilbourne2006influenza,shaman2010absolute,moore2012improvement,kissler2020projecting,stone2007seasonal}}, heterogeneity in community susceptibility \cite{glass2004effect,van2001measles}, and targeted vaccination \cite{ball2006optimal}---can also be evaluated through appropriate modifications to $\tilde{P}^{*}$ and $\tilde{P}_{\mathrm{th},ij}$. For example, targeted vaccination yielding perfect immunity is typically implemented by removing nodes from the network \cite{ball2006optimal,wang2016statistical}; in our framework, this could equivalently be accomplished by setting $P_{\mathrm{th},ij}>P^{*}$, therefore preventing further infection of a node, while imperfect immunization \cite{mclean1993imperfect, halloran1989modeling} could alternatively be implemented through smaller increases in $P_{\mathrm{th},ij}$.


\section*{Methods}


\noindent\textbf{Implementation of the network model.} We implement the dynamic network model in MATLAB. To define each 2D square lattice of 100 by 100 individuals (``nodes''), we specify node locations and an adjacency matrix characterizing the connectivity of the network. To ensure only one horizontally-oriented border between the two subpopulations in the stratified population shown in Fig. \ref{SI Fig2}, we employ a periodic boundary condition in the horizontal direction by connecting the first and last nodes in each row for all simulations; the nodes at the top and bottom boundaries do not have such conditions, ensuring that the vertical direction is non-periodic. For the bonds between nodes (``edges''), we randomly assign the values of the interaction barrier $\tilde{P}_{\mathrm{th},ij}\in[0,1]$ from a given distribution, as specified in the main text. Each simulation has a specified disease infectivity $\tilde{P}^{*}$ and recovery duration $\tilde{\tau}_{r}$. From the values of $\tilde{P}_{\mathrm{th},ij}$ and $\tilde{P}^{*}$, we compute the discrete infection transmission times $\Delta \tilde{\tau}_{i j}=1 / (1-\tilde{P}_{\mathrm{th}, i j} / \tilde{P}^{*})$ for each edge. The simulations shown in Figs. \ref{fig1}-\ref{fig2}, \ref{fig3}a,d, and \ref{fig4}a are for the exact same lattice with the exact same configuration of $\tilde{P}_{\mathrm{th},ij}$ taken from a uniform distribution. Similarly, the simulations shown in Figs. \ref{fig3}b,e, and \ref{fig4}b are for the exact same lattice with the exact same configuration of $\tilde{P}_{\mathrm{th},ij}$ taken from a normal distribution, and the simulations shown in Figs. \ref{fig3}c,f, and \ref{fig4}c are for the exact same lattice with the exact same configuration of $\tilde{P}_{\mathrm{th},ij}$ taken from a bimodal distribution. 

To perform each simulation, we use a modified invasion percolation algorithm based on the method described by Masson \textit{et al.} \cite{masson2014fast}. We start at $\tilde{\tau}=0$ by introducing the disease at the central node of the lattice, with all the other nodes specified as being susceptible ($S$). Then, for the next and each successive time step of the simulation, we use a binary tree structure to sort all edges in contact with infected nodes and find the most susceptible edge $ij$---the edge with the minimal infection transmission time $\Delta \tilde{\tau}_{i j}$. The next node to become infected, $j$, is then the node that is connected to the infected region through this most susceptible edge. This target node is then specified as being infected ($I$), and its time of infection is specified by adding the time increment $\Delta\tilde{\tau}_{i j}$ to the overall elapsed time $\tilde{\tau}$. We also decrease the remaining transmission times for all edges in contact with infected nodes by this time increment. New edges made available for infection by node $j$ are added to the binary tree; because the tree was mostly sorted in the last step, subsequent sorts are time-efficient. We incorporate recovery by identifying all infected nodes for which at least $\tilde{\tau}_{r}$ has elapsed since infection, setting its infection transmission time to all nodes connected to it as being equal to $\infty$, and specifying the node as being recovered ($R$). \\





\section*{Supplementary Information} 

\begin{figure}
    \centering
    \includegraphics[width=3in]{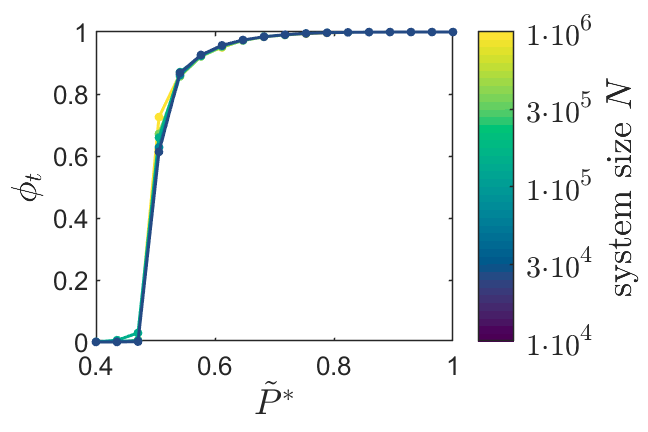}
    \caption{Critical growth of the total infected fraction $\phi_t$ above a critical infectivity $\tilde{P}^*>0.5$ for a recovery-free population. Network size is varied logarithmically from $N=10^4$ to $10^6$ with no discernible difference (some curves lie beneath $N=10^4$).}
    \label{fig:sizeDep}
\end{figure}

\noindent\textbf{Size dependence.} Our simulations occur on a fixed finite size network. All simulations presented are conducted on networks with $N=10^4$ nodes. We verify that our results are not considerably influenced by finite-size  effects by repeating simulations for five network sizes. Figure \ref{fig:sizeDep} shows that the critical growth of the total infected fraction $\phi_t$ above $\tilde{P}^*>0.5$, corresponding to Fig. 1d in the main text, is insensitive the system size, even when the number of nodes is increased by two orders of magnitude ($N=10^4$ to $10^6$). We hence anticipate our system size $N=10^4$ to be sufficiently large to capture the general dynamics of this spread.\\

\noindent\textbf{Estimate of characteristic disease spreading timescales.} For the case of a recovery-free population, we calculate the shortest possible time $\tilde{\tau}_{f,0}$ at which the infected fraction plateaus in the limit of high disease infectivity $\tilde{P}^*$. As time progresses, the disease spreads radially outward. Because we consider a square network comprising $N_{t}$ nodes in total, $\sqrt{N_{t}}$ on a side, the leading edge of the circular infected region first reaches the boundary of the population when $\tilde{\tau}\approx\sqrt{N_{t}}/2$. However, the total infected fraction can continue to grow: it only plateaus when it spans the entire 2D network, including its corners. This occurs when the radius of the infected region is equal to half the diagonal of the square network, $\tilde{\tau}_{f,0}\approx(\sqrt{N_{t}}/2)\times\sqrt{2}=\sqrt{N_{t}/2}$. 


For the case of a population with recovery duration $\tilde{\tau}_{r}$, we extend this calculation to estimate the time at which the infected fraction of the population will peak, $\tilde{\tau}_p$, as well as the time at which the infected fraction of the population reaches zero after all individuals recover, $\tilde{\tau}_f$, again in the limit of high disease infectivity. As time progresses, the disease spreads radially outward in a circular infected region, followed by an inner circular region of recovery that spreads at the same rate but is delayed by $\tilde{\tau}_{r}$. We consider two separate regimes: the ``thin pulse'' regime with $\tilde{\tau}_{r}<\sqrt{N_{t}}(1/\sqrt{2}-1/2)$, and the ``thick pulse'' regime with $\tilde{\tau}_{r}>\sqrt{N_{t}}(1/\sqrt{2}-1/2)$.

\begin{figure}
\centering
\includegraphics[width=3.5in]{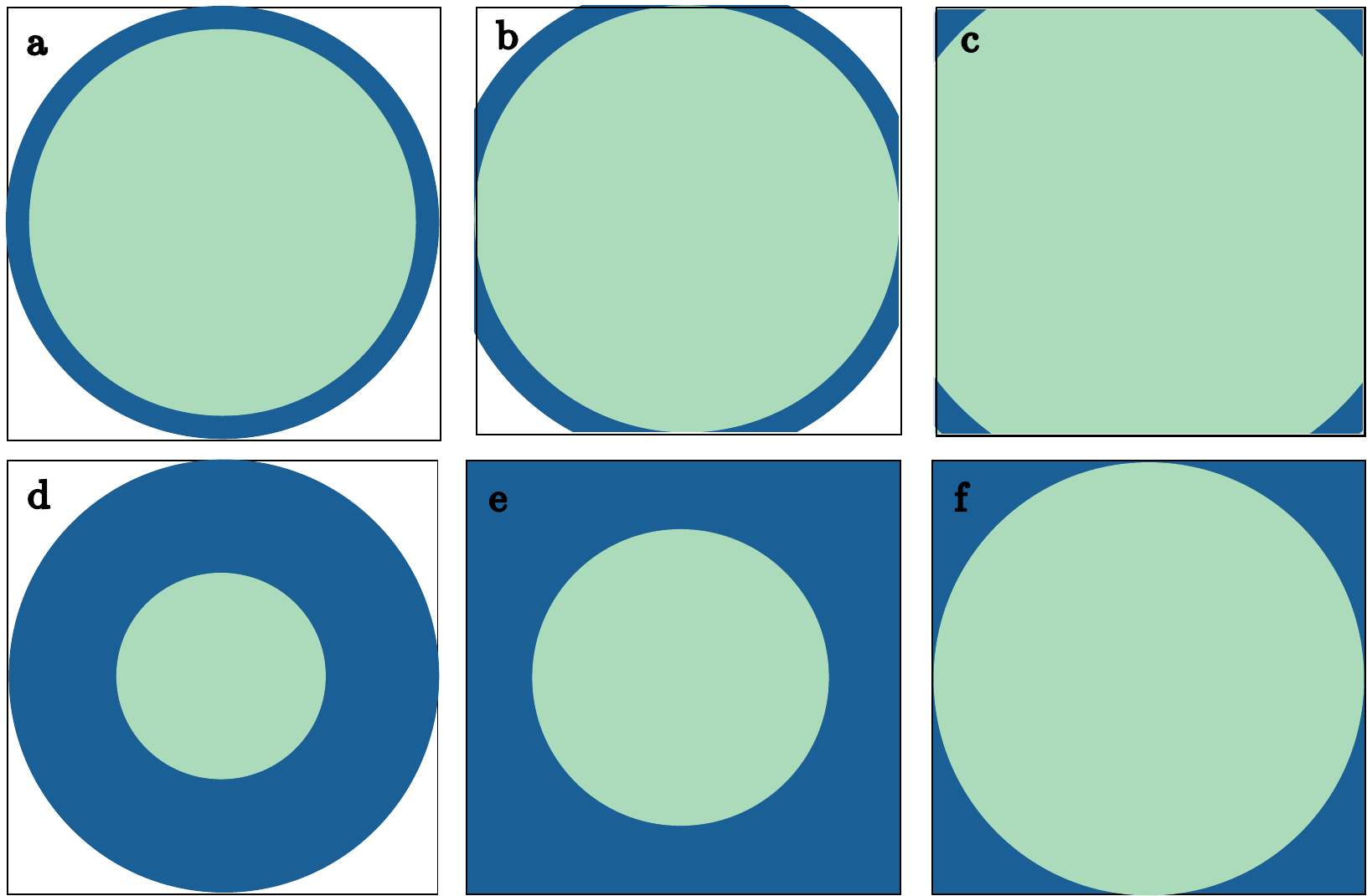}
\caption{\label{SI Fig1} Schematics showing the growth of the regions of infection (dark blue) and recovery (light green) over time. Top row shows the thin pulse regime with low $\tilde{\tau}_{r}$ while bottom row shows the thick pulse regime with high $\tilde{\tau}_{r}$. In the thin pulse regime, \textbf{(a)} the leading edge of the infected population reaches the boundary first, \textbf{(b)} followed by the leading edge of the recovered population, \textbf{(c)} followed by the leading edge of the region of infection circumscribing the entire population. However, in the thick pulse regime, \textbf{(d)} while the leading edge of the infected population again reaches the boundary first, \textbf{(e)} the region of infection reaches the corners of the square lattice before \textbf{(g)} the leading edge of the recovered population reaches the boundary.}
\end{figure}

For a thin pulse, as in the recovery-free case, the leading edge of the infected region first reaches the boundary of the population when $\tilde{\tau}\approx\sqrt{N_{t}}/2$ (Fig. \ref{SI Fig1}a). At this time, the total infected fraction is nearly maximal, and we therefore approximate $\tilde{\tau}_{p}\approx\sqrt{N_{t}}/2$. As time progresses, the leading edge of the region of recovery then first reaches the boundary of the population at a time $\tilde{\tau}\approx\sqrt{N_{t}}/2+\tilde{\tau}_{r}$ (Fig. \ref{SI Fig1}b). Both regions continue to spread into the corners of the square boundary, and the leading edge of the infected region eventually reaches the corners at a time $\tilde{\tau}\approx\sqrt{N_{t}/2}$ as in the recovery-free case (Fig. \ref{SI Fig1}c). Subsequently, the region of recovery continues to grow; the total infected fraction continues to decrease, eventually reaching zero when the region of recovery has reached the corners of the square boundary, $\tilde{\tau}_{f}\approx\sqrt{N_{t}/2}+\tilde{\tau}_{r}$.

For a thick pulse, the leading edge of the infected region again first reaches the boundary of the population when $\tilde{\tau}\approx\sqrt{N_{t}}/2$ (Fig. \ref{SI Fig1}d). The leading edge of the infected region then reaches the corners of the square boundary at a time $\tilde{\tau}\approx\sqrt{N_{t}/2}$ as in the recovery-free case (Fig. \ref{SI Fig1}e). Thus, the time at which the infected fraction is maximal is between these two times: $\sqrt{N_{t}}/2\lesssim\tilde{\tau}_{p}\lesssim\sqrt{N_{t}/2}$. As time progresses, the region of recovery then continues to grow, eventually first reaching the boundary of the population at $\tilde{\tau}\approx\sqrt{N_{t}}/2+\tilde{\tau}_{r}$ (Fig. \ref{SI Fig1}f). Subsequently, the region of recovery continues to grow; the total infected fraction continues to decrease, eventually reaching zero when the region of recovery has reached the corners of the square boundary, $\tilde{\tau}_{f}\approx\sqrt{N_{t}/2}+\tilde{\tau}_{r}$.

For our simulations with $N_t = 10^4$, the transition between the thin and thick pulse regimes occurs at $\tilde{\tau}_r \approx 21$; therefore, our analysis of the example system with $\tilde{\tau}_r =4$ presented in the main text is in the thin pulse regime, with $\tilde{\tau}_{p}\approx\sqrt{N_{t}}/2\approx50$ and $\tilde{\tau}_{f}\approx\sqrt{N_{t}/2}+\tilde{\tau}_{r}\approx75$ as reported in the main text. Together with Eq. 2, these estimates provide a universal scaling for the peak infection time, $\phi_p = \phi(\tilde{\tau}_p)$ (Figs. 3d-f insets). 

\section*{Supporting movie captions} 

\noindent\textbf{Movie S1.} Sequence of infection for a disease with low infectivity $\tilde{P}^{*}=0.3$, showing that disease spreading is quickly localized. This simulation is without recovery.\\

\noindent\textbf{Movie S2.} Sequence of infection for a disease with intermediate infectivity $\tilde{P}^{*}=0.6$, showing that the disease spreads in a spatially heterogeneous, ramified pattern, leading to the formation of discrete clusters of bypassed individuals who remain uninfected. Infected individuals are shown in dark blue, uninfected susceptible individuals are shown in white. This simulation is without recovery.\\


\noindent\textbf{Movie S3.} Sequence of infection for a disease with higher infectivity $\tilde{P}^{*}=0.7$, showing that the disease spreads in a more compact pattern, with a smoother leading edge, leading to the formation of fewer and smaller clusters of bypassed individuals. Infected individuals are shown in dark blue, uninfected susceptible individuals are shown in white. This simulation is without recovery.\\


\noindent\textbf{Movie S4.} Sequence of infection for a disease with intermediate infectivity $\tilde{P}^{*}=0.6$, showing that recovery causes disease spreading to be quickly localized. Infected individuals are shown in dark blue, recovered individuals are shown in light green, uninfected susceptible individuals are shown in white. This simulation is with $\tilde{\tau}_{r}=4$.\\


\noindent\textbf{Movie S5.} Sequence of infection for a disease with higher infectivity $\tilde{P}^{*}=0.7$, showing that the disease spreads continually, but recovery causes the disease to spread in a spatially heterogeneous, ramified pattern, leading to the formation of discrete clusters of bypassed individuals who remain uninfected. Infected individuals are shown in dark blue, recovered individuals are shown in light green, uninfected susceptible individuals are shown in white. This simulation is with $\tilde{\tau}_{r}=4$.

\begin{acknowledgments}
It is a pleasure to acknowledge Navid C.P.D. Bizmark for stimulating discussions. This work was supported by startup funds from Princeton University. This material is also based upon work supported by the National Science
Foundation Graduate Research Fellowship Program (to C.A.B.) under Grant No. DGE1656466. Any opinions, findings, and conclusions or recommendations expressed in this
material are those of the authors and do not necessarily reflect the views of the National
Science Foundation.\\

\noindent \textbf{Author contributions}: C.A.B., D.B.A., and J.S. initiated the project; D.B.A. designed simulations; C.A.B. and J.S. performed all simulations; S.S.D. designed and performed the theoretical analysis; All authors designed the overall project, analyzed the data, discussed the results and implications, and wrote the manuscript; and S.S.D. supervised the overall project.\\

\noindent\textbf{Competing Interests}: The authors declare no competing interests.
\end{acknowledgments}

\providecommand{\noopsort}[1]{}\providecommand{\singleletter}[1]{#1}%


\end{document}